\documentclass[letterpaper,prl,twocolumn,superscriptaddress,showpacs]{revtex4-1} 

\usepackage{graphicx, color}

\begin{document}


\title{Perspective: Heusler interfaces -- opportunities beyond spintronics?}

\author{Jason K. Kawasaki}
	\email{jkawasaki@wisc.edu}
	\affiliation{Department of Materials Science and Engineering, University of Wisconsin, Madison WI 53706} 

 	
\date{\today}

\begin{abstract}

Heusler compounds, in both cubic and hexagonal polymorphs, exhibit a remarkable range of electronic, magnetic, elastic, and topological properties, rivaling that of the transition metal oxides. To date, research on these quantum materials has focused primarily on bulk magnetic and thermoelectric properties or on applications in spintronics. More broadly, however, Heuslers provide a platform for discovery and manipulation of emergent properties at well-defined crystalline interfaces. Here, motivated by advances in the epitaxial growth of layered Heusler heterostructures, I present a vision for Heusler interfaces, focusing on the frontiers and challenges that lie beyond spintronics. The ability to grow these materials epitaxially on technologically important semiconductor substrates, such as GaAs, Ge, and Si, provides a direct path for their integration with modern electronics. Further advances will require new methods to control the stoichiometry and defects to ``electronic grade'' quality, and to control the interface abruptness and ordering at the atomic scale.

\end{abstract}

\maketitle

The properties at materials interfaces often exceed the simple sum of their bulk constituents. Prime examples include the two-dimensional electron gas at the interface between insulators LaAlO$_3$ and SrTiO$_3$ \cite{ohtomo2004high}; the order of magnitude enhancement of the superconducting critical temperature at the monolayer FeSe/SrTiO$_3$ interface \cite{qing2012interface, he2013phase, lee2014interfacial}; topological states at the interfaces between topological materials and normal materials (or a surface) \cite{hsieh2008topological, chen2009experimental}; modulation doping in semiconductor heterostructures \cite{dingle1978electron} for discoveries in fundamental physics (e.g. integer \cite{klitzing1980new, tsukazaki2007quantum} and fractional \cite{tsui1982two, stormer1999fractional} quantum Hall effects) and application in high electron mobility transistors (HEMTs \cite{dingle1979high, asif1993high, mimura2002early}), and the band bending, carrier confinement, and current rectification at semiconductor interfaces (e.g. GaAs/AlGaAs, Si/Ge) \cite{dingle1978electron, kroemer1957theory}, which form the backbone of the modern microelectronics industry. Today, Herb Kroemer's insight that ``the interface is the device'' extends well beyond the originally envisioned semiconductor interfaces \cite{kroemer2001nobel}.

In searching for new interfacial materials platforms, it is desirable to not only identify parent materials with a wide range of new functionality, but also to integrate these materials epitaxially with technologically important substrates. Heusler compounds are promising in both regards. The ternary intermetallic Heusler compounds have long held promise for their bulk thermoelectric \cite{mastronardi1999antimonides, zeier2016engineering} and magnetic properties \cite{graf2011simple, wollmann2017heusler}, especially half-metallic ferromagnetism at room temperature for applications in spintronics \cite{de1983new}. But their functional properties extend well beyond these topics of early focus, and now include topological states \citep{chadov2010tunable, lin2010half, logan2016observation, liu2016observation, manna2018heusler}, superconductivity with novel pairing \cite{brydon2016pairing, kim2018beyond, wernick1983superconductivity, nakajima2015topological, timm2017inflated, wang2018unconventional}, (ferro, antiferro, and ferri)-magnetism \cite{suzuki2016large, tobola, sanvito2017accelerated}, skyrmions \cite{nayak2017magnetic, meshcheriakova2014large}, superelasticity and shape memory effect (ferroeleasticity) \cite{liu2003martensitic}, and predicted ferroelectricity and \textit{hyper}ferroelectricity \cite{garrity2014hyperferroelectrics,bennett2012hexagonal}. Combining these functionalities at atomically defined interfaces presents opportunities to discover and manipulate emergent properties that do not exist in the bulk, e.g., via the application of epitaxial strain, quantum confinement, proximity effects, and band discontinuities. Furthermore, due to their close symmetry- and lattice-match to zincblende semiconductors, many Heuslers can be grown epitaxially on semiconductor substrates such as GaAs, GaSb, Si, and Ge, providing a template for their integration with modern electronics.

In this Perspective I offer a vision for Heusler interfaces as a platform for interfacial materials discovery and design. A number of excellent reviews on magnetic Heuslers and their applications in spintronics already exist, for which the reader is referred to Refs. \cite{farshchi2013spin, felser2015basics, hirohata2006heusler, wollmann2017heusler, palmstrom2016heusler, palmstrom2003epitaxial, casper2012half, felserbook, graf2011simple}. Here I look ahead to the opportunities and challenges that lie beyond spintronics, where the plethora of competing ground states \cite{chadov2010tunable, lin2010half, manna2018heusler}, combined with advances in epitaxial film growth \cite{bach2003molecular, palmstrom2016heusler, kawasaki_cotisb}, theory \cite{kawasaki2018simple, sharan2019formation, picozzi2007polarization}, and characterization \cite{kawasaki2018simple, logan2016observation, liu2016observation, brown2018epitaxial, Andrieu_co2mnsi, jourdan2014direct}, position Heuslers as an intriguing platform for interfacial materials design. Further advances will require new methods to control the stoichiometry and defects in Heuslers to ``electronic grade'' quality, and to control interface abruptness and ordering at the atomic scale. I discuss the opportunities and challenges for Heusler interfaces, in both cubic and hexagonal polymorphs. The broad array of functionality in Heuslers is highly complementary to that of the transition metal oxides \cite{mannhart2010oxide}, but from the opposite starting point: whereas oxides are generally brittle insulating ceramics that can be made conductive, Heuslers are ductile conductive intermetallics that can be made insulating.

\begin{figure*}
 \includegraphics[width=7in]{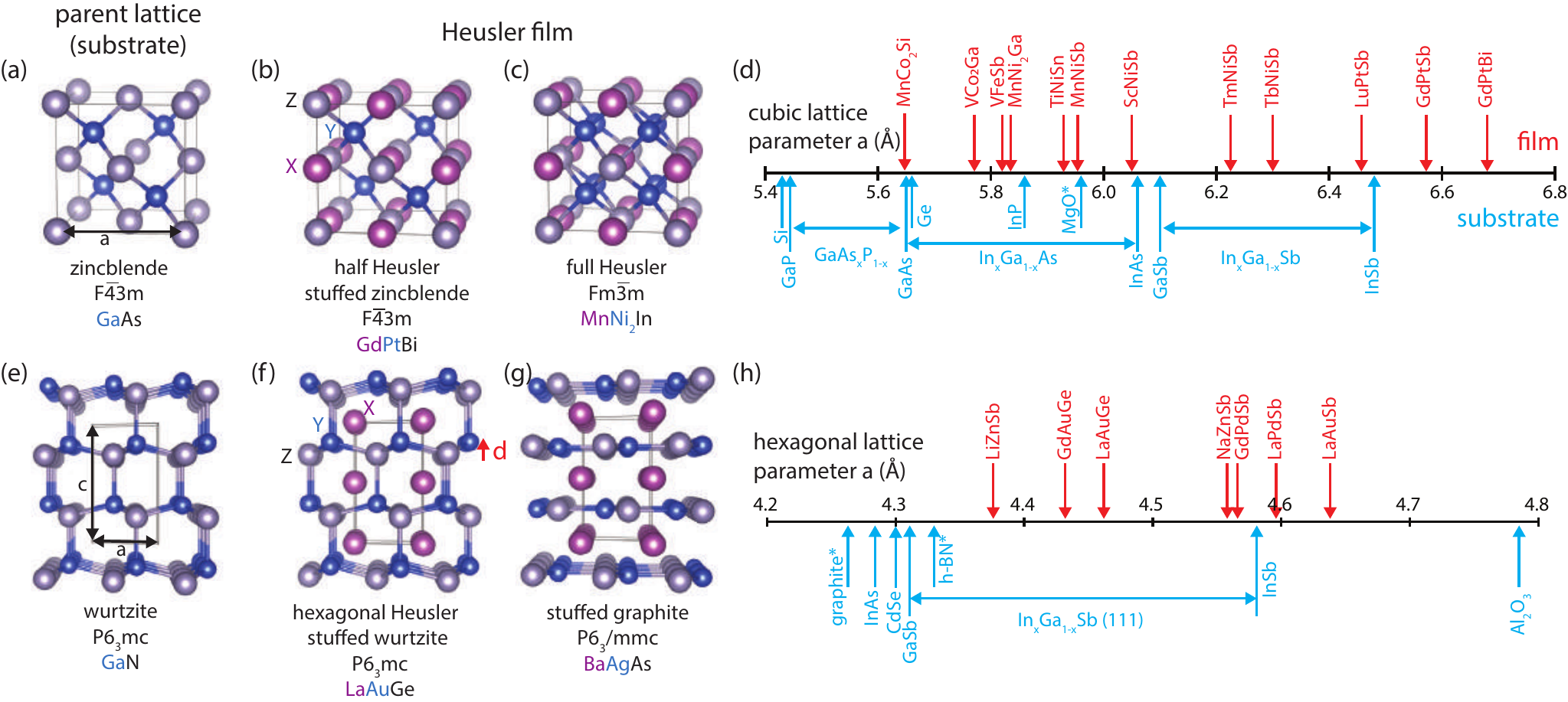}
 \caption{Crystal structures for cubic and hexagonal Heuslers, (a-c) Cubic Heuslers compared to the parent zincblende lattice. (b) The half Heusler contains an additional sublattice of $X$ at $(\frac{1}{2}, 0, 0)$. (c) Full Heuslers contain an additional $Y$ sublattice at $(\frac{3}{4}, \frac{1}{4}, \frac{1}{4})$. (e-g) Hexagonal Heusler compared to the parent wurtzite lattice. (f) The hexagonal \textit{LiGaGe}-type ``stuffed wurtzite'' structure is characterized by a polar buckling of the $YZ$ planes. (g) The \textit{ZrBeSi}-type ``stuffed graphite'' structure can be viewed as ``stuffed graphite.'' (d,h) Lattice parameters for cubic and hexagonal Heuslers, and comparison to commercially available substrates and epitaxial buffer layers. The asterisks (*) denote a rotation of 45 (cubic) or 30 (hexagonal) degrees around the [001] or [0001] axis. Hexagonal lattice parameters for the zincblende III-V semiconductors (InAs, GaSb, InSb) refer to $\{ 111 \}$ orientation.}
 \label{crystal}
\end{figure*}

\section*{Quantum materials, minus the oxygen}
Heusler compounds are intermetallic compounds that crystallize in one of several stuffed tetrahedral structures. Conventionally the term ``Heusler'' has been reserved for the cubic polymorphs, but here I expand the term to include the hexagonal analogues (Fig. \ref{crystal}). The cubic half Heusler compounds (Fig. \ref{crystal}b, spacegroup $F\bar{4}3m$, $C1_b$ structure) have composition $XYZ$ and consist of a zincblende $[YZ]^{n-}$ sublattice that is ``stuffed'' with $X^{n+}$ at the octahedrally coordinated sites $(\frac{1}{2}, 0, 0)$ \cite{kandpal}. Alternatively, this structure can be viewed as a rocksalt $XZ$ sublattice that is ``stuffed'' with $Y$ in every other tetrahedral interstitial site $(\frac{1}{4}, \frac{1}{4}, \frac{1}{4})$ \cite{ougut1995band,larson2000structural}. Full Heusler compounds, with composition $X Y_2 Z$, contain an additional $Y$ atom in the basis to fill all of the tetrahedral interstitials (Fig. \ref{crystal}c, $L2_1$ structure) \cite{graf2011simple}. Here I adopt the naming convention of ordering the elements by \textit{increasing} electronegativity, i.e. $XYZ$ and $X Y_2 Z$ (TiCoSb and MnNi$_2$Ga ), rather than the often adopted $YXZ$ and $Y_2 X Z$ (CoTiSb and Ni$_2$MnGa)\cite{knowlton1912heusler, heusler1903magnetische}, for consistency with standard naming conventions of other compounds
\footnote{Multiple conventions have been used for naming cubic and hexagonal Heusler compounds. The naming of hexagonal Heuslers typically follows the convention of listing constituents in order of increasing electronegativity $XYZ$, e.g. LaCuSn and LiGaGe. The naming of cubic half Heusler compounds often follows this $XYZ$ convention, e.g. LuPtBi, TiCoSb, and TiNiSn; however, the ordering $YXZ$ is also commonly used, e.g. CoTiSb and NiTiSn. This alternate convention exists because full Heusler compounds are more commonly written $Y_2 XZ$ (e.g., Ni$_2$MnGa) rather than $XY_2 Z$ (MnNi$_2$Ga). For consistency across the cubic and hexagonal Heusler compounds, this article adopts the electronegativity convention. When another notation is more commonly used, that name is also mentioned.}.
Lattice parameters for most cubic Heuslers are spanned by the zincblende III-V semiconductors, from GaP ($5.45$ \AA) to GaAs ($5.653$ \AA) and InSb ($6.479$ \AA).  Relaxed ternary buffer layers, e.g., In$_x$Ga$_{1-x}$As, enable the lattice parameter to be tuned exactly (Fig. \ref{crystal}d). 

\begin{figure*}
 \includegraphics[width=7in]{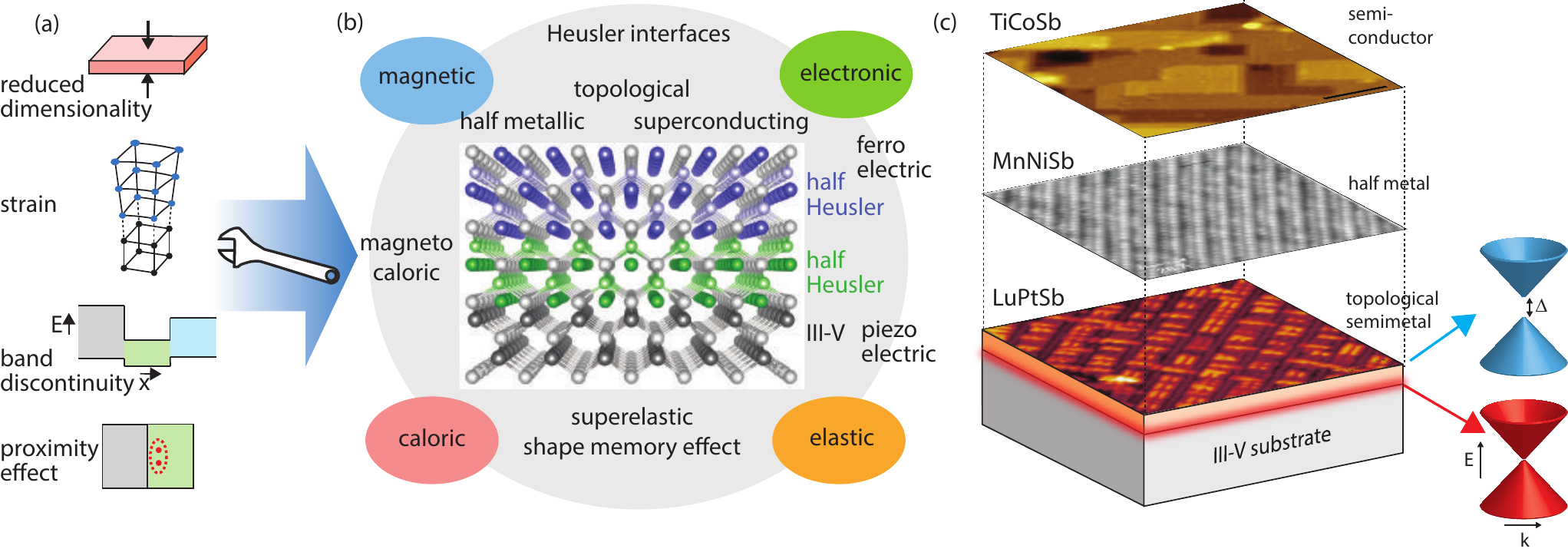}
 \caption{Emergent properties at Heusler interfaces. (a) Tuning parameters for epitaxial films and heterostructures. (b) Heusler properties, grouped into four mator functionalities: magnetic, electronic, elastic, and caloric. (c) Scanning tunneling microscopy (STM) images of TiCoSb (semiconductor), MnNiSb (half metallic ferromagnet), and LuPtSb (topological semimetal). Each of these compounds can be grown on a III-V substrate. A Dirac dispersion of topological states is expected at the interface between topological semimetals and normal III-V substrates. A gapped Dirac cone and quantum anomalous Hall effect are expected at the interface between topological materials and ferromagnets. TiCoSb STM adapted from J. K. Kawasaki \textit{et. al.}, Sci. Adv. 4: eaar5832 (2018) \cite{kawasaki2018simple}. Reprinted with permission from AAAS. MnNiSb STM reprinted figure with permission from P. Turban, \textit{et. al.,} Phys. Rev. B 65, 134417 (2002) \cite{turban}. Copyright 2002 by the American Physical Society. LuPtSb STM courtesy of N. Wilson and C. Palmstr{\o}m.}
 \label{props}
\end{figure*}

Hexagonal polymorphs also exist, in which polar distortions give rise to properties that are not symmetry-allowed for cubic systems, e.g., ferroelectricity. The parent \textit{ZrBeSi}-type structure is nonpolar and consists of planar graphite-like $[YZ]^{n-}$ layers that are ``stuffed'' with $X^{n+}$ (spacegroup $P6_3 / mmc$ Fig. \ref{crystal}g). Unidirectional buckling $d$ of the $[YZ]^{n-}$ layers produces the polar \textit{LiGaGe}-type structure, which can be viewed as a ``stuffed wurtzite'' (spacegroup $P6_3 mc$, Fig. \ref{crystal}f) \cite{casper2008searching, hoffmann2001alb2}. Many insulating $P6_3 mc$ materials are promising as ferroelectrics, with calculated polarization and energy barrier to switching comparable to BaTiO$_3$ \cite{bennett2012hexagonal, garrity2014hyperferroelectrics}, despite being composed of all-metallic  constituents. These predictions challenge the conventional notion that good ferroelectrics should be highly ionic materials with large Born effective charges \cite{benedek2014polarization}. In fact, it is precisely this lack of ionicity and stronger tendency towards covalent bonding that is predicted to make many hexagonal Heuslers robust against the depolarizing field, making them \textit{hyperferroelectrics} \cite{garrity2014hyperferroelectrics}. Other polar $P6_3 mc$ materials are natively semimetallic, and are of interest as low resistivity polar metals \cite{benedek2016ferroelectric}. These hexagonal materials are ``stuffed'' versions of wurtzite GaN (polar) and hexagonal BN (nonpolar), and can be lattice matched to zincblende semiconductor substrates in $\{ 111 \}$ orientation (Fig. \ref{crystal}h). 

In both hexagonal and cubic polymorphs, $X$ is typically a transition or rare earth metal, $Y$ a transition metal, and $Z$ a main group metal (III, IV, or V). The phase boundary between hexagonal and cubic polymorphs is determined by the relative size of the $X$ cation compared to $Y$ and $Z$, with larger $X$ cations favoring the hexagonal polymorphs and smaller $X$ favoring cubic polymorphs \cite{seibel2015gold, hoffmann2001alb2, casper2008searching, xie2014pressure}. This dependence on chemical pressure suggests that the phase boundary could also be traversed by epitaxial strain, giving the epitaxial grower access to phases that would otherwise be challenging to stabilize and retain by bulk synthetic methods, e.g., hydrostatic pressure. Other structural variants include Jahn-Teller driven cubic to tetragonal distortions, variations in the atomic site ordering (e.g. ``inverse Heusler,'' $D0_3$, and $B2$ cubic variants), and polar vs antipolar layer buckling patterns in the hexagonal variants \cite{seibel2015gold, bennett2013orthorhombic, strohbeen2019electronically}.


The wide array of quantum properties in these materials arises from the large orbital degeneracy and the spatial confinement of $d$ and $f$ orbitals (compared to $s$ and $p$), with rich phenomena that are highly complementary to that observed in the transition metal oxides. Due to their lack of oxygen or other highly electronegative species, Heusler compounds are typically less ionic than oxides and the on-site electron-electron repulsion (Coulomb $U$) is generally weaker. For these reasons, Heusler compounds are unlikely to be a good system for finding new high temperature superconductors \cite{dagotto1994correlated}. On the other hand, magnetic exchange interactions (Hund's $J$) are quite significant, as evidenced by the strong tendency for magnetic ordering with Curie temperatures as large as 1100 K \cite{wurmehl2006investigation}. The lack of oxygen combined with substrate lattice matching make Heuslers more amenable than oxides to integration with compound semiconductors, since oxygen interdiffusion, reactions, and misfit dislocations pose significant challenges for oxide on semiconductor epitaxy \cite{hubbard1996thermodynamic}. Additionally, many Heusler compounds can be alloyed via to form quaternary, quinary, and even higher component alloys, providing a means to continuously tune the lattice parameter and properties of the Heusler compound itself \cite{zeier2016engineering}. Finally, whereas oxides are typically brittle, Heuslers are highly elastic, displaying superelasticity and accommodating strains of several percent without plastic deformation \cite{bungaro2003first, dong2004shape}, making them attractive for flexible magnetoelectronics.

\begin{figure*}
 \includegraphics[width=7in]{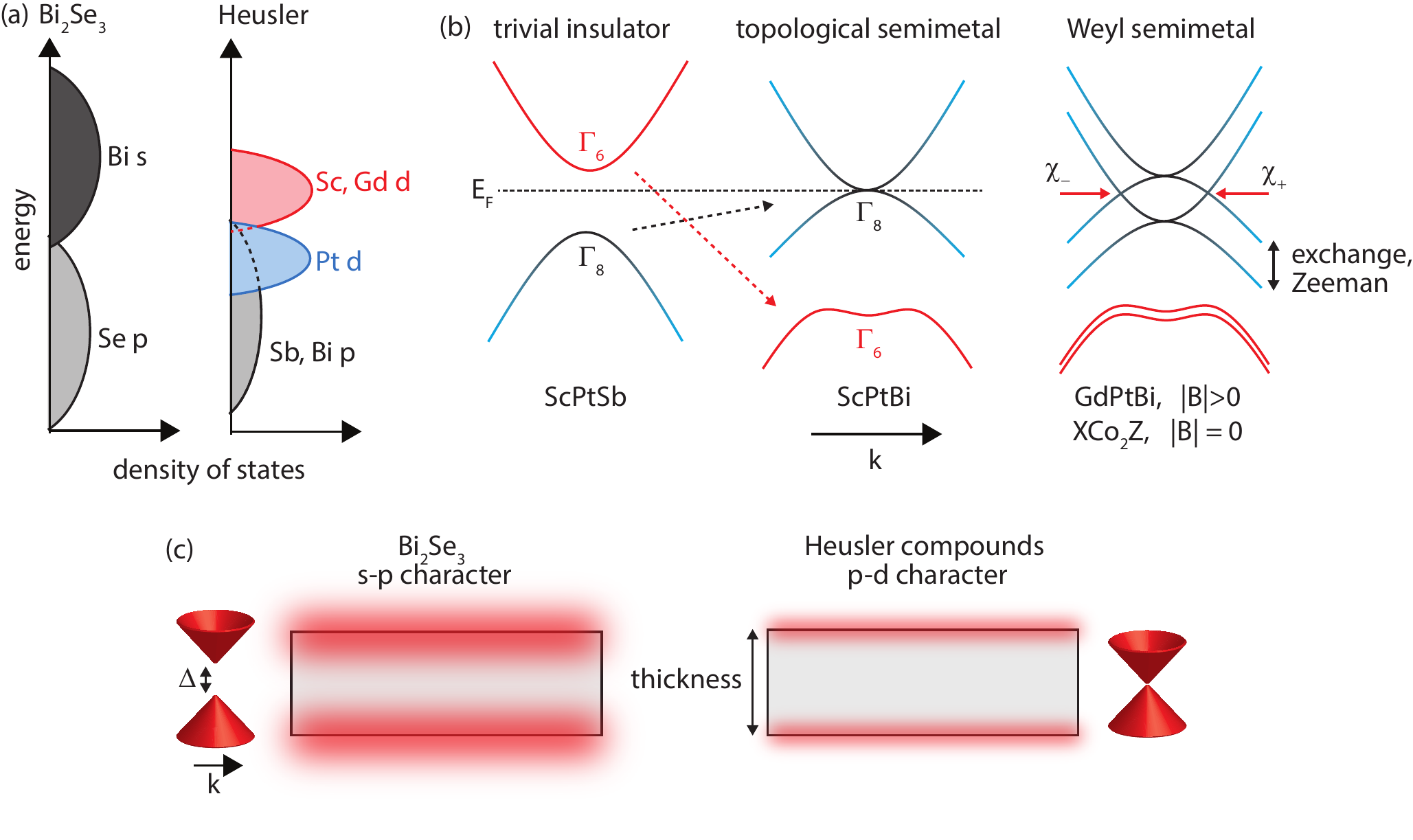}
 \caption{Topological states in $d$-band Heusler compounds. (a) Schematic density of states for Bi$_2$Se$_3$ and Heusler compounds ScPtSb, ScPtBi, and GdPtBi. Heusler compounds show significant $d$ character from the $X$ (Sc, Gd) and $Y$ (Pt) rare earth and transition metal sites. (b) Schematic energy-momentum dispersions. Starting from the trivial insulator ScPtSb (left), substitution of Sb with the heavy Bi atom leads to a spin-orbit induced band inversion of the $\Gamma_6$ and $\Gamma_8$ bands, creating a topological semimetal with quadratic band touchings (middle). GdPtBi (right), which has partially filled $f^7$ levels, is antiferromagnetic in its ground state \cite{nakajima2015topological}. Upon the application of an external magnetic field, the combination of Zeeman splitting and the exchange field creates a pair of linearly dispersing Weyl nodes with opposite chirality $\chi_{+}$ and $\chi_{-}$ \cite{hirschberger2016chiral}. For $X$Co$_{2}Z$ compounds, which are ferromagnetic in the ground state, Weyl nodes are expected in the absence of an applied magnetic field \cite{wang2016time}. (c) Schematic of topological state hybridization for ultrathin films. Due to the smaller spatial extent of $d$ orbitals than $s$ and $p$ orbitals, the critical thickness for gap opening in Heuslers is expected to be smaller than for Bi$_2$Se$_3$.}
 \label{topological}
\end{figure*}

\section*{Opportunities at interfaces}

Many of the most intriguing properties arise when the diverse properties of Heuslers can be combined and manipulated at atomically defined interfaces (Fig. \ref{props}). Such interfaces include Heusler/semiconductor, Heusler/oxide, and interfaces between two different Heusler compounds. Here I highlight several opportunities that lie beyond spintronics, and in the next section I describe the key experimental challenges.

\textbf{Topological states.} Recent angle-resolved photoemission spectroscopy (ARPES) measurements \cite{logan2016observation, liu2016observation} confirm theoretical predictions \cite{lin2010half, chadov2010tunable} of topological surface and interface states in cubic half Heuslers with large spin-orbit coupling compared to the bandwidth ($\lambda_{SO} > W$), e.g. $R$PtBi and $R$PtSb ($R=$ rare earth metal) (Fig. \ref{topological}b). Such states arise at the interfaces between topologically band-inverted materials and normal materials (or the vacuum, i.e. a surface), and are of great importance for dissipationless transport and for discovery of emergent quasiparticles when interfaced with layers of other functionality, e.g., Majorana bound states at topological / superconductor interfaces \cite{fu2008superconducting}.

Compared to first-generation binary topological insulators Bi$_x$Sb$_{1-x}$ and Bi$_2$Se$_3$, Heuslers offer several distinct opportunities. Firstly, whereas in most known topological materials the near-Fermi level states have $s$ and $p$ character, Heuslers have significant transition metal $d$ character with moderate electron-electron correlations \cite{chadov2009electron} (Fig. \ref{topological}a). The interplay between correlations and spin-orbit coupling is predicted to yield rich correlated topological properties in other systems, e.g. axion states and topological Kondo insulators in iridates. \cite{kim2008novel, witczak2014correlated}. Heuslers provide an alternative materials system for realizing such phenomena. Another potential consequence of localized $d$ orbitals is a shorter critical length scale for surface and interface state hybridization (Fig. \ref{topological}c). ARPES measurements of ultrathin Bi$_2$Se$_3$ films reveal that below a critical thickness of six quintuple layers ($\sim 6$ nm), the topological states at top and bottom interfaces hybridize to open a gap \cite{zhang2010crossover}. The smaller spatial extent of $d$ states in Heuslers implies that topological states may survive to smaller critical thicknesses without gapping out.

Secondly, the multifunctionality within the Heusler family enables lattice-matched topological heterostructures, for interfacing topological states with layers of other functionality. For example, topological / superconductor interfaces are predicted to host Majorana bound states, and topological / ferromagnet interfaces are expected to exhibit the quantum anomalous Hall effect \cite{chang2013experimental, liu2016quantum} and axion states \cite{xiao2018realization, mogi2017magnetic}. Lattice matching minimizes the potentially detrimental effects of misfit dislocations and interfacial defect states that could otherwise obscure the property of interest \cite{richardella2015characterizing}, e.g., by acting as parasitic conduction channels.

Finally, Heuslers are a platform for other topological states, including Dirac and Weyl fermions, in both cubic and hexagonal polymorphs. In cubic Heuslers, transport signatures of Weyl nodes have been observed in several $R$PtBi compounds \cite{hirschberger2016chiral, liang2018experimental, shekhar2018anomalous, PhysRevB.99.035110} and MnCo$_2$Ga (also known as Co$_2$MnGa) \cite{sakai2018giant} under an applied magnetic field, and theory predicts Weyl nodes in the magnetic full Heuslers $X$Co$_{2}Z$ without an external field \cite{wang2016time} (Fig. \ref{topological}b). In the hexagonal polymorphs, which break inversion symmetry, DFT calculations predict Weyl nodes \cite{narayan2015class, gao2018dirac} whose  momenta are highly sensitive to the magnitude of polar buckling, potentially tunable by epitaxial strain.

\textbf{Interfacial superconductivity.} Heuslers are a platform for novel superconductivity, both at artificially defined interfaces and natively due to strong spin-orbit coupling. Whereas most known superconductors have singlet pairing, triplet superconductivity is predicted at interfaces between conventional superconductors $(S)$ and ferromagnets $(F)$ \cite{bergeret2001long}. Signatures of triplet pairing have been observed experimentally in Heusler-based $S/F/S$ Josephson junctions, where $S=$ Nb and $F=$ MnCu$_2$Al \cite{sprungmann2010evidence}. All-Heusler Josephson junctions offer the potential of realizing such behavior in fully-lattice matched systems that minimize interfacial disorder. Examples of Heusler superconductors include the cubic full Heuslers $X$Pd$_2$Sn ($X = $Sc, Y, Lu) and $X$Pd$_2 Z$ [$X=$ (Zr, Hf), $Z=$ (In, Al)] (also known as Pd$_{2}XZ$) \cite{klimczuk2012superconductivity, winterlik2009superconductivity}; the rare earth containing half Heuslers $R$PdBi \cite{nakajima2015topological, kim2018beyond}; and the hexagonal compounds BaPtAs \cite{kudo2018superconductivity}, SrPtAs \cite{nishikubo2011superconductivity}, and YbGaSi \cite{imai2008superconductivity}. Heusler $S/F/S$ Josephson junctions are also a platform $\pi$ phase control in an all epitaxial system, with potential applications as qubits \cite{yamashita2005superconducting}.

Intriguingly, recent theory \cite{brydon2016pairing, venderbos2018pairing} and experiments \cite{kim2018beyond} suggest triplet and higher order pairing may exist \textit{natively} in a subset of topological superconducting half Heuslers with composition $R$PdBi. Here the pairing occurs between $j=3/2$ fermions due to strong spin-orbit coupling. This combination is expected to natively host Majorana states in a single material \cite{yang2017majorana}, in contrast with previous experimental realizations that rely on an interface between a superconductor and a separate high spin-orbit material \cite{mourik2012signatures}.

\textbf{Interface polarization: ferroelectrics and polar metals.} For conventional ferroelectrics, the depolarizing field typically competes with and destroys long range polar order in the limit of ultrathin films. Hexagonal Heusler interfaces offer two potential solutions to this problem. Firstly, a number of insulating $P6_3 mc$ compounds (e.g., LiZnAs, NaZnSb) have been proposed as \textit{hyperferroelectrics}, which are robust against the depolarizing field due to their highly covalent bonding character with small Born effective charges \cite{garrity2014hyperferroelectrics}. Ferroelectric switching and hyperferroelectricity have yet to be experimentally demonstrated in hexagonal Heuslers. A significant challenge is that only a small subset of hexagonal Heuslers is natively insulating - an assumed requirement for switching via applied electric fields \cite{fei2018ferroelectric}. Epitaxial strain, quantum confinement, and Peierls-like distortions \cite{seibel2015gold, strohbeen2019electronically, genser2019coupling} may provide routes for tuning the buckling and opening a gap in polar compounds that are natively metallic.

For those hexagonal compounds that cannot be made insulating, the coexistence of a polar structure and metallicity holds interest in its own right, and may be a second solution to the depolarizing field problem. Polar metals, once assumed to be unstable since free carriers were thought to screen out polar displacements, are not fundamentally forbidden \cite{anderson1965symmetry} and have recently been demonstrated in several transition metal oxides \cite{shi2013ferroelectric, cao2018artificial, kim2016polar, kim2016polar}. Hexagonal $P6_3 mc$ Heuslers are another family of polar metals, and are unique in that they are generally more conductive than oxides \cite{kaczorowski1999magnetic, schnelle1997crystal}. One application for polar metals may be to suppress the effects of the depolarizing field by pinning displacements at the polar metal / ferroelectric interface \cite{puggioni2016polar}. Other opportunities for polar metal interfaces may lie in nonlinear optics \cite{wu2017giant}, nonlinear charge transport \cite{kang2019nonlinear, tokura2018nonreciprocal}, and novel superconductivity \cite{edelstein1995magnetoelectric}. 

\begin{figure}[h]
 \includegraphics[width=3.5in]{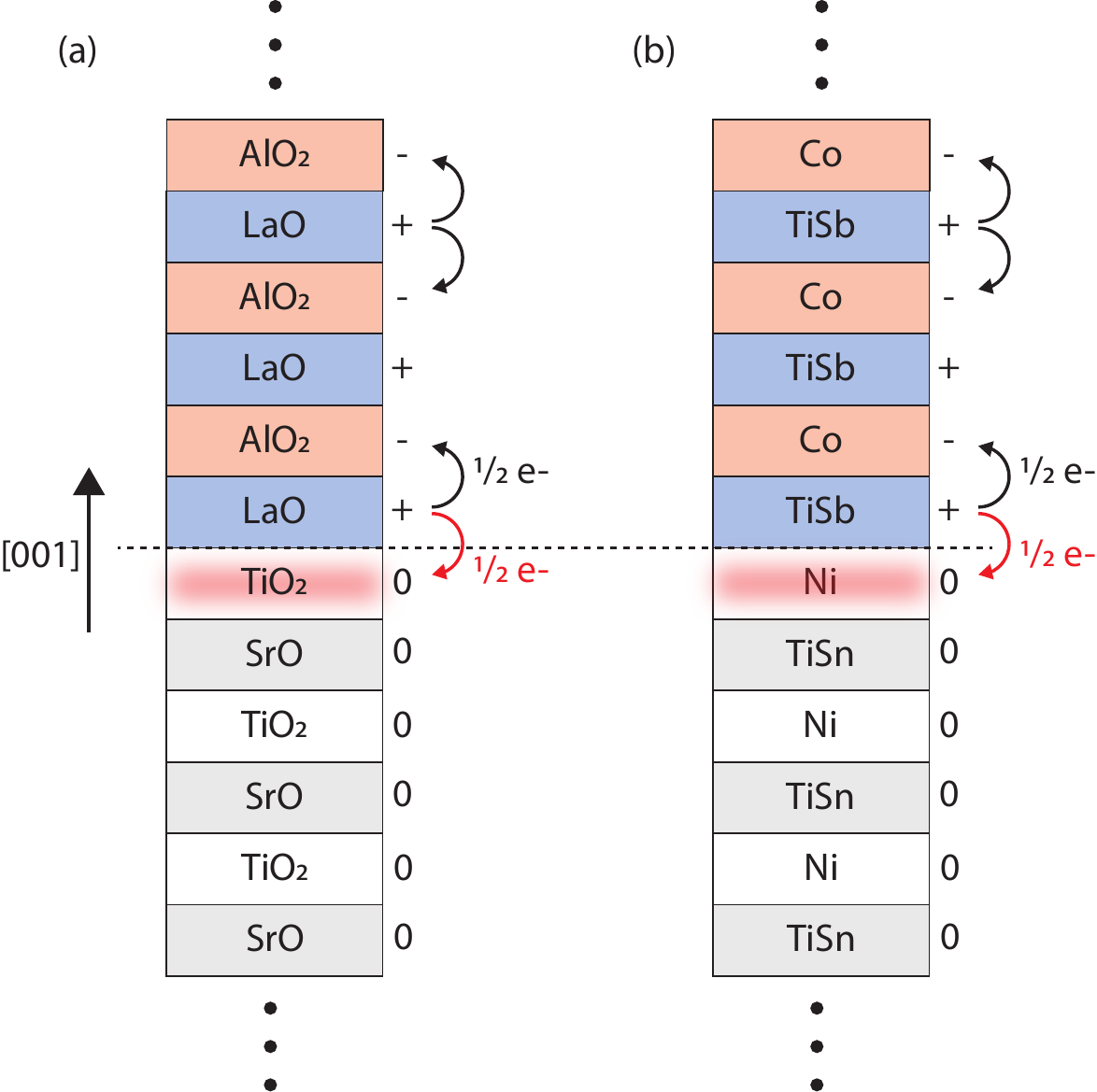}
 \caption{Mechanism for polar catastrophe 2DEG at oxide and Heusler interfaces. (a) In bulk LaAlO$_3$, each LaO layer donates half an electron per formula unit to the AlO$_2$ layer above and to the AlO$_2$ layer below, resulting in alternating formal charges of $+1/-1$. This results in an excess charge of half an electron per formula unit at the (001) oriented interface with SrTiO$_3$, which consists of charge neutral SrO and TiO$_2$ planes. (b) A similar mechanism is expected for the half Heusler system TiCoSb / TiNiSn, in which TiCoSb is composed of alternating charged planes while TiNiSn is composed of charge neutral planes.}
 \label{polarcatastrophe}
\end{figure}

\textbf{Polar catastrophe.} Interface polarization and charge transfer also provide opportunities for the creation of two-dimensional electron gasses (2DEGs) across interfaces. Consider the classic ``polar catastrophe'' 2DEG that emerges at the LaAlO$_3$ / SrTiO$_3$ interface (Fig. \ref{polarcatastrophe}a). In this $3d$ electron system, the (001) stacking sequence of SrTiO$_3$ consists of charge neutral SrO/TiO$_2$ atomic planes, while LaAlO$_3$ consists of LaO/AlO$_2$ atomic planes with alternating $+1 / -1$ charge \cite{ohtomo2004high, janotti2012controlling}. The 2DEG arises from charge transfer of half an electron per formula unit across the interface, from the LaO atomic plane in LaAlO$_3$ to the TiO$_2$ atomic plane in SrTiO$_3$ \cite{ohtomo2004high, janotti2012controlling}. The half Heusler system TiNiSn / TiCoSb contains the same essential ingredients: just like LAO/STO, the near Fermi level orbitals also have strong $3d$ character. In (001) orientation the Ni/TiSn atomic planes are formally charge neutral, while the Co/TiSb planes have formal charges $-1 / +1$ \cite{kawasaki2018simple} (Fig. \ref{polarcatastrophe}b). These formal charges are based on an electron counting model \cite{kawasaki2018simple} that accurately predicts the experimentally measured surface reconstructions of TiCoSb (001), and is consistent with the experimental data for LuPtSb \cite{patel2014surface, patel2016surface2}, MnNiSb \cite{bach2003molecular}, and TiNiSn (001) \cite{kawasaki_nitisn}.

In this highly simplified view, a charge transfer 2DEG might also be expected at this Heusler interface \cite{sharan2019formation}. Recent transport measurements on MBE-grown TiCoSb/TiNiSn bilayers show 1.5 order of magnitude enhanced conductivity \cite{ricethesis}, consistent with this interfacial charge transfer prediction \cite{sharan2019formation}. Additionally, in LaAlO$_3$ / SrTiO$_3$ the strong spatial confinement of the 2DEG has been suggested to enhance electron-electron correlations and contribute to the emergent superconductivity. What new properties may emerge at Heusler interfaces with enhanced correlations?


\textbf{Interfacial magnetism and skyrmions.} Heusler interfaces offer a platform for enhancing skyrmion phase stability via combined bulk and interfacial inversion symmetry breaking. Magnetic skyrmions are topologically protected vortex-like swirls of spin texture, whose robustness against atomic scale disorder makes them attractive for applications in magnetic memory. They are stabilized \cite{roessler2006spontaneous, binz2006theory, han2010skyrmion} by the Dzyaloshinskii-Moriya (DM) exchange interaction \cite{dzyaloshinsky1958thermodynamic, moriya1960anisotropic}, which results from a combination of broken inversion symmetry and large spin-orbit coupling. To date, most work has focused on two separate strategies to stabilize skyrmions: (1) bulk crystal structures that break inversion, e.g. $B20$ crystals such as FeGe and MnSi \cite{muhlbauer2009skyrmion}, or (2) artificially defined interfaces that break inversion \cite{bogdanov2001chiral}, e.g. Co/Pt interfaces \cite{yang2015anatomy}. 

Combining bulk and interfacial DM interactions in a single materials platform is predicted to be a path towards further control and enhancements of skyrmion stability \cite{rowland2016skyrmions}. Heuslers are a strong materials candidate. Recent experiments confirm skyrmions in several Mn$_2 YZ$ compounds that crystallize in the tetragonal inverse Heusler structure ($I\bar{4}m2$) that breaks bulk inversion \cite{nayak2017magnetic, meshcheriakova2014large}. The epitaxial film growth of several of these compounds has recently been demonstrated \cite{jin2018structural, swekis2019topological, rana2016observation, meshcheriakova2015structural, li2018large} providing a path towards further manipulation of the DM interaction in layered heterostuctures of these materials. Beyond skyrmion stability, recent theoretical proposals suggest that skyrmion/superconductor interfaces may be another platform for hosting Majorana fermions \cite{PhysRevB.93.224505, PhysRevB.92.214502}, potentially realizable in an all-Heusler system.

\begin{figure}[th]
 \includegraphics[width=3.4in]{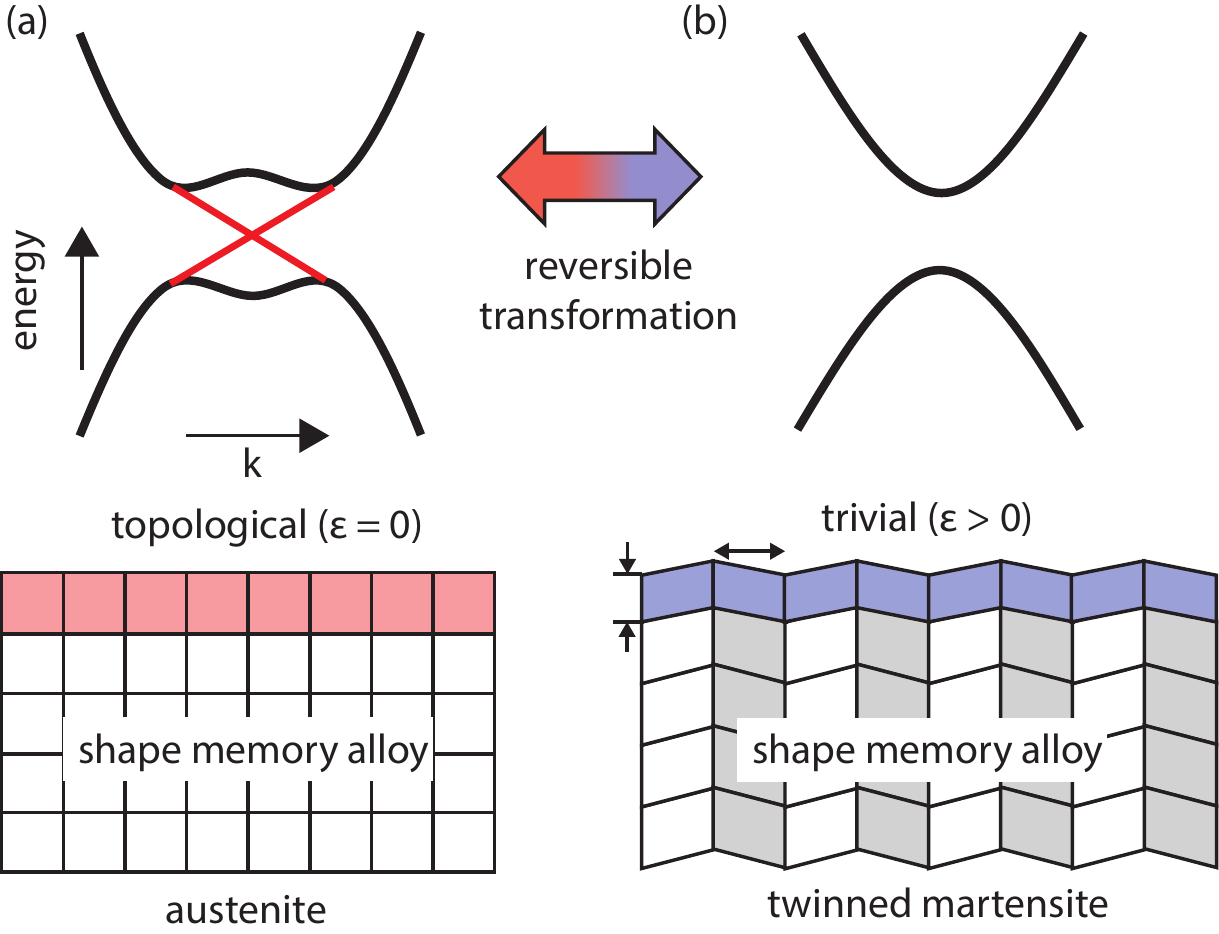}
 \caption{Concept for a topological switch, induced by reversible martensitic phase transitions. The shape memory alloy undergoes a displacive transformation from the high symmetry austenite phase to a low symmetry twinned martensite, as a function of temperature or applied magnetic field. Strains across the interface induce a structural distortion in the ultrathin Heusler layer, e.g. $R$PtBi, transforming it from a topological phase to a trivial phase.}
 \label{martensite}
\end{figure}

\textbf{Interface strain and shape memory effect.} Shape memory alloys are ferroelastic materials that undergo large, reversible martensitic phase transitions or twin reorientations to revert a macroscopically deformed material back to its original shape. Several Heuslers, including MnNi$_2 Z$ ($Z=$ group III or IV) and MnCo$_2$Ga, exhibit such transitions, driven by temperature and strain (shape memory effect), or by an external magnetic field (magnetic shape memory effect) \cite{dong2004shape, bungaro2003first}. These compounds are also known as Ni$_2$Mn$Z$ and Co$_2$MnGa. Across these transitions the magnetic, caloric, and electrical transport properties change abruptly \cite{manosa2010giant}, and these materials are generally also superelastic, accommodating strains as large as 10 percent by locally undergoing strain-induced martensitic phase transitions or twin reorientations \cite{sozinov2002giant}. This 10 percent strain is an order of magnitude larger than the strains observed in magnetostrictive or piezoelectric materials, with promising applications for microactuation and vibration dampening. The large latent heat associated with the phase transition holds promise for applications in refrigeration and thermal energy conversion \cite{srivastava2011direct, song2013thermodynamics}.

Layered heterostructures composed of a shape memory alloy provide an opportunity to couple the large and reversible strains across materials interfaces, to \textit{induce} phase transitions in adjacent functional layers. For example, DFT calculations suggest that the topological band inversion in the $R$PtBi Heuslers can be flipped by strains of approximately 3\% \cite{chadov2010tunable}. One could envision $R$PtBi / shape memory alloy interfaces in which the topological states are switched ``on'' and ``off'' by temperature or magnetic field-induced martensitic phase transitions (Fig. \ref{martensite}). Strains of this magnitude are likely too large to be produced by coupling to magnetostrictive or piezoelectric layers, but are within the limits of shape memory alloys.

Well-defined epitaxial interfaces also provide an idealized test bed for understanding and manipulating the phase transition itself. A key limiting factor in bulk shape memory alloys is that the habit plane, i.e. the interface between austenite and martensite phases, is not guaranteed to be atomically commensurate (epitaxial). As a result, repeated cycling through the martensitic phase transition creates dislocations that lead to slower switching speeds, decreased energy efficiency, and eventually mechanical failure \cite{gall2002cyclic}. A promising materials design route is to engineer materials such that the habit plane is atomically commensurate or near-commensurate, i.e. the \textit{compatibility} and \textit{cofactor} conditions \cite{james2000martensitic, james1989theory, bhattacharya1991wedge, ball100i987, gu2018phase}. This condition is met when the middle eigenvalue $\lambda_2$ of the austenite to martensite transformation matrix equals 1. One design route towards the $\lambda_2=1$ criterion is to deliberately fabricate non-stoichiometric samples \cite{cui2006combinatorial, chluba2015ultralow}, such as Mn$_{25+y}$Ni$_{50-x}$Co$_x$Sn$_{25-y}$ (also known as Ni$_{50-x}$Co$_x$Mn$_{25+y}$Sn$_{25-y}$) \cite{srivastava2010hysteresis, bhatti2012small}.

Another route is to engineer the habit plane via film/substrate interface effects in epitaxial thin films, which can be tuned via crystallographic orientation and strain \cite{bhattacharya1999tents, kaufmann2011modulated}. For example, for epitaxial NiTi films grown on (001) oriented MgO \cite{buschbeck2011martensite} and GaAs \cite{buschbeck2011growth} substrates, clamping effects from the substrate force a new transformation pathway in which the habit plane lies parallel to the (001) \cite{buschbeck2011martensite}. Importantly, this transformation occurs via a shear mechanism in which the interface remains atomically coherent, and may provide a general route towards engineering atomically commensurate phase transitions.

\section*{Challenges}


Significant advances have been made on the epitaxial growth and control of Heusler interfaces over the past 20 years, primarily driven by applications in spintronics. These include the development of Heusler molecular beam epitaxy (MBE) \cite{ambrose2000epitaxial, van2000epitaxial, bach2003molecular, dong1999molecular, dong2000epitaxial, turban}, the identification of semi adsorption-controlled growth windows \cite{kawasaki2018simple, patel2014surface, bach, turban, strohbeen2019electronically}, the use of epitaxial diffusion barriers and low temperature seed layers \cite{palmstrom2016heusler, farshchi2013spin, schultz2002eras, buschbeck2011growth}, the use of chemical templating layers \cite{dong2000epitaxial, filippou2018chiral}, and the development simple theoretical frameworks based on electron counting \cite{kandpal, jung2000study, pashley1989electron} for predicting stability and structural distortions at surfaces and interfaces \cite{kawasaki2018simple, pashley1989electron}.

Despite these advances, the full realization of Heusler properties beyond spintronics will likely require even more stingent control of materials and interface quality. This is because many of the emerging properties in Heuslers depend on bandgaps: bulk bandgaps in topological insulators and ferroelectrics, minority spin gaps in half metals, and pairing gaps in superconductors. Such gaps tend to be highly sensitive to non-stoichiometry, point defects, lattice distortions, and interfacial reconstructions and disorder. Additionally, interfacial properties are often inherently short-range, and therefore can be sensitive subtle changes in atomic structure across the interface.

\begin{figure}[h]
 \includegraphics[width=2.6in]{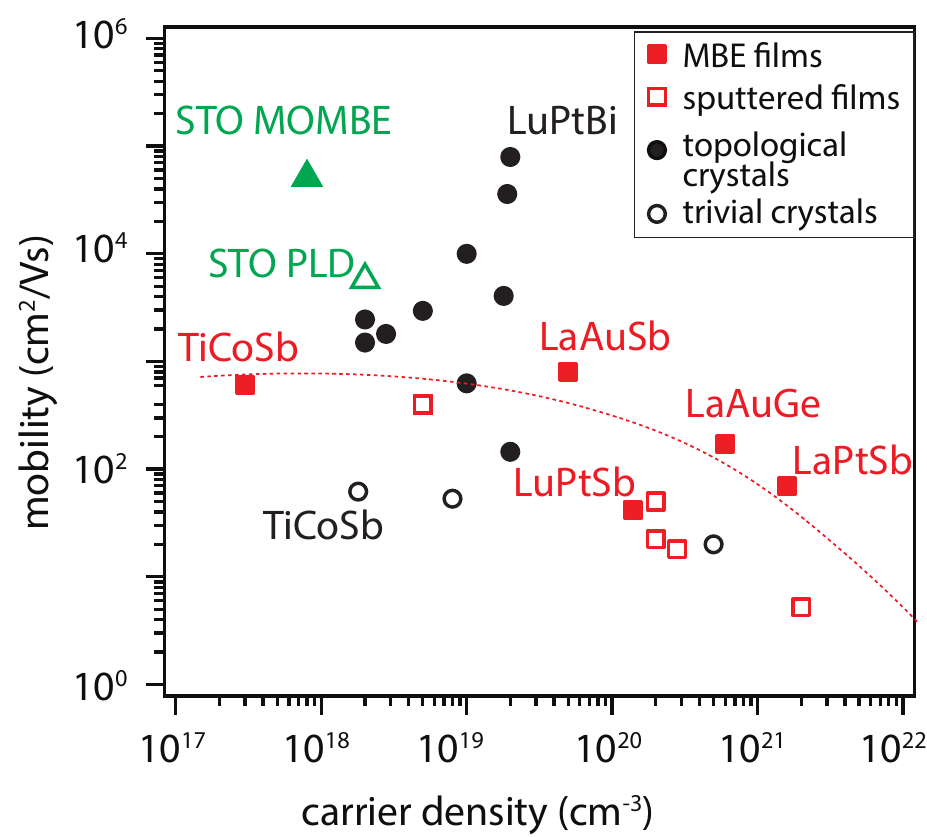}
 \caption{Carrier mobility and density at 2K for 18 valence electron half Heuslers, in bulk crystal and epitaxial film form. Legend: MBE-grown films (filled red squares \cite{strohbeen2019electronically, kawasaki_cotisb, patel2014surface,kawasaki_thesis}), sputter-grown films (open red squares, Refs. \cite{jaeger2011epitaxial, wang2012fabrication, narita2015effect, shan2013electronic}), single crystal topological semimetals $R$(Pt,Bd)(Sb,Bi) (filled black circles, Refs.\cite{hou2015high, hirschberger2016chiral, hou2015large, nakajima2015topological}), and bulk semiconductors (open black circles, Refs. \cite{ahilan2004magnetotransport, wu2007thermoelectric}). For the MBE-grown samples, TiCoSb was grown on lattice matched InAlAs/InP(001) \cite{kawasaki_cotisb} and on MgO(001) \cite{kawasaki_thesis}, LuPtSb on InAlSb/GaSb(001) \cite{patel2014surface}, and the hexagonal compounds LaAuSb, LaAuGe, and LaPtSb on Al$_2$O$_3$(0001) \cite{strohbeen2019electronically}. Most sputtered films were grown on MgO(001). For comparison, I also show the oxide SrTiO$_3$, grown by pulsed laser deposition (PLD \cite{kozuka2010dramatic}) and by adsorption-controlled MOMBE \cite{cain2013doped}.}
 \label{transport}
\end{figure}

\textbf{Controlling stoichiometry and defects to ``electronic grade.''} Although bandstructure calculations predict a number of half Heuslers to be semiconductors with bandgaps of 1 eV or larger, typical background carrier densities are well above $10^{17}$ cm$^{-3}$ and mobilities below $500$ cm$^2$/Vs, for both bulk crystals and thin films (Fig. \ref{transport}). Flux-grown single crystals of Heusler topological semimetals do have higher mobilities approaching $10^5$ cm$^2$/Vs (filled black circles \cite{hou2015high, hirschberger2016chiral, hou2015large, nakajima2015topological}); however, this higher mobility results in part from the topological protection of surface or bulk Dirac and Weyl states rather than purely a reduction of bulk impurity scattering.


\begin{figure}[th]
 \includegraphics[width=2.5in]{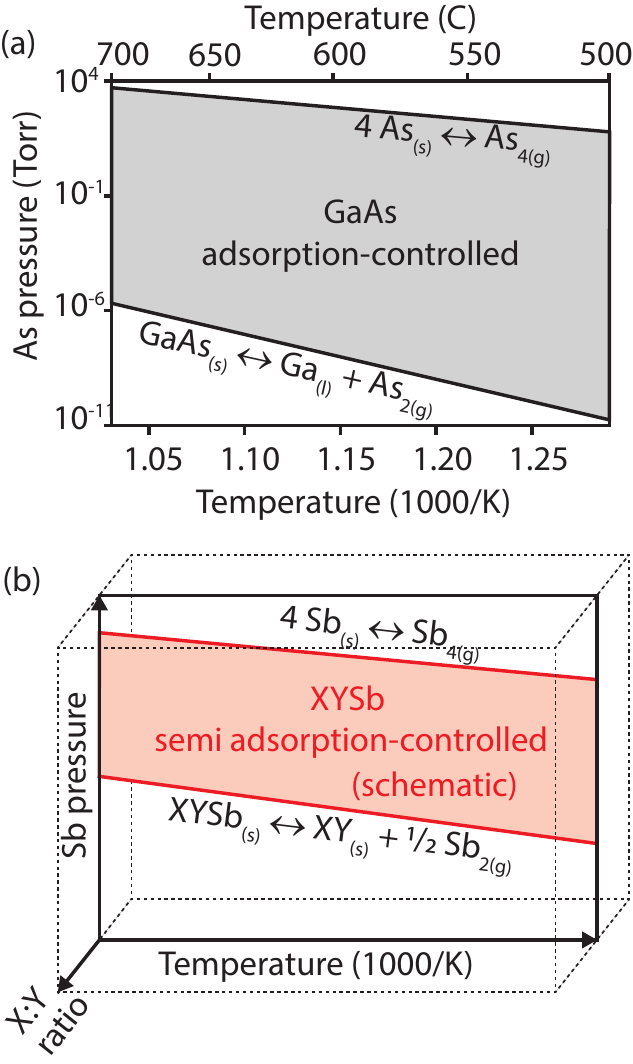}
 \caption{Thermodynamics of adsorption-controlled growth. (a) Growth window for stoichiometric GaAs, as a function of arsenic partial pressure and sample temperature. Adapted from Ref. \cite{theis1998adsorption}. The bounds of this growth window are determined by the vaporization of arsenic (upper bound) and the decomposition of GaAs into Ga liquid and As$_2$ vapor (lower bound). Within the window, stoichiometric solid GaAs plus As$_2$ vapor is formed. (b) Schematic semi-adsorption-controlled window for antimonide Heuslers. The upper bound is given by the vaporization of Sb, while the lower bound is given by the decomposition of the Heusler phase. One possible decomposition reaction, $XY$Sb$_{(s)}$ $\leftrightarrow XY_{(s)} +$ $\frac{1}{2}$Sb$_{2(g)}$, is shown. The Sb stoichiometry is self limiting; however, the transition metal stoichiometry $X:Y$ is not.}
 \label{adsorption}
\end{figure}

The poor transport properties stem largely from challenges in controlling the stoichiometry and resultant defects, which are generally more difficult to control in ternary intermetallics than in simple binary semiconductors. To illustrate this challenge, consider binary GaAs, which shows record high electron mobility when grown in a modulation doped structure by MBE \cite{dingle1978electron, pfeiffer2003role, gardner2016modified}. A major reason for the success of MBE-grown GaAs is the existence of a thermodynamically adsorption-controlled growth window \cite{tsao2012materials}, in which the stoichiometry is self-limiting (Fig. \ref{adsorption}(a)). Due to the high volatility of arsenic, GaAs films are grown with an excess arsenic flux, in which only the stoichiometric As:Ga ratio ``sticks'' and the excess arsenic escapes in the vapor phase. High mobility ternary III-V alloys, e.g. In$_x$Ga$_{1-x}$As, are also routinely grown by MBE in which the As:(In+Ga) stoichiometry is self-limiting. The In:Ga stoichiometry is not self-limiting; however, since both In and Ga have the same valence and incorporate substitutionally on the same lattice sites, slight variations of In:Ga composition result in subtle changes in the bandgap rather than the formation of defect states.

In select cases, ternary Heuslers can be grown in a \textit{semi} adsorption-controlled window (Fig. \ref{adsorption}b), in which the stoichiometry of \textit{one} of the three elements is self-limiting. TiCoSb \cite{kawasaki2018simple, kawasaki_cotisb}, MnNiSb \cite{bach, turban}, LuPtSb \cite{patel2014surface}, and LaAuSb \cite{strohbeen2019electronically} can be grown by MBE with an excess Sb flux, in which the ratio of Sb to $(X+Y)$ is self-limiting.The TiCoSb films grown by this method display the lowest background carrier concentration ($\rho(300K) = 9 \times 10^{17}$ cm$^{-3}$, $\rho(2K) = 2 \times 10^{17}$ cm$^{-3}$) of any gapped Heusler compound to date \cite{kawasaki_cotisb}, including bulk crystals (Fig. \ref{transport}). The electron mobility of $530$ cm$^2$/Vs is similarly large, compared to typical values of less 100 cm$^2$/Vs for growth by sputtering or arc melting (Fig. \ref{transport}). For the semimetal LaAuSb grown by semi adsorption-controlled MBE, the 2K mobility is 800 cm$^2$/Vs \cite{strohbeen2019electronically}.

However, it remains an outstanding challenge to control the remaining $X:Y$ transition metal stoichiometry. This is especially important for Heuslers, compared to III-V ternary alloys, since $X$ and $Y$ occupy different lattice sites and typically have different valences. At typical growth temperatures of $300-600^{\circ}$C the sticking coefficients for elemental transition metals are near unity, therefore the film $X:Y$ stoichiometry relies on precise control of $X$ and $Y$ fluxes rather than a self-limiting process. Due to typical flux drifts, these fluxes are difficult to control to better than $1\%$, even when using real-time flux monitoring and feedback approaches such as optical atomic absorption \cite{chalmers1993real, kasai1997atomic}) or x-ray emission (RHEED-TRAXS \cite{hasegawa1985chemical}) spectroscopies. In a worst-case scenario, if all nonstoichiometric defects were electrically active, a $1\%$ deviation in stoichiometry would correspond to an unintentional carrier density of order $10^{20}-10^{21}$ cm$^{-3}$, clearly unacceptable for most electronic applications. At such high concentrations, the defects typical form an impurity band or a ``perturbed host'' band \cite{yonggang2017natural}. While not all defects are electronically active \cite{yonggang2017natural}, experimentally it is found that most polycrystalline half Heuslers have carrier densities greater than $10^{20}$ cm$^{-3}$ \cite{muta2009high, kim2007high, fu2015realizing}. In general only flux-grown single crystals and semi adsorption-controlled MBE films have unintentional densities below $10^{20}$ cm$^{-3}$ (Fig. \ref{transport}). Control of stoichiometry is also critical for half-metallic ferromagnets, since non-stoichiometric defects often produce states within the minority spin gap \cite{picozzi2007polarization}, thus decreasing the spin polarization at the Fermi energy.

One possible solution may to replace one or both of the transition metal sources with a volatile chemical precursor. For transition metal oxides, fully adsorption-controlled growth of SrTiO$_3$ and SrVO$_3$ thin films has been demonstrated by replacing elemental Ti and V with titanium tetra-isopropoxide (TTIP) and vanadium tetra-isopropoxide (VTIP), respectively. The resulting films exhibit record high electron mobility \cite{cain2013doped, son2010epitaxial} and low residual resistivity \cite{moyer2013highly}, exceeding their bulk counterparts. 
This approach is generally called metalorganic molecular beam epitaxy (MOMBE) \cite{putz1985gaas} or chemical beam epitaxy (CBE) \cite{tsang1984chemical}. First developed in the 1980's for growth of III-Vs, MOMBE was applied a few years later to the growth of superconducting oxides $R$Ba$_2$Cu$_3$O$_{7-x}$ ($R =$ Y, Dy) \cite{king1991situ, endo1991preparation}. For the case of perovskite oxides, this approach has recently been termed \textit{hybrid} MBE (\textit{h}MBE) \cite{jalan2009molecular, jalan2009growth}, where the distinction \textit{hybrid} refers to the combined use of metalorganic$+$elemental$+$gas sources \cite{brahlek2018frontiers}, as opposed to purely metalorganic or metalorganic$+$elemental sources. Given the remarkable success of volatile precursor MBE for transition metal oxide growth, similar advances are anticipated if the approach can be applied to Heuslers. Potential precursors include metalorganics, e.g. the metal cyclopentadienyls, or volatile metal halides. However, such precursors introduce new challenges of potential carbon incorporation and equipment corrosion, respectively.

Ultimately, the degree of stoichiometry control possible by adsorption-control may be limited by the phase diagram of the particular system. For example, rather than existing as pure line compounds, some Heusler compounds have a finite (few percent) phase field along certain directions in composition space. The $x<0.05$ solubility of excess Ni within TiNi$_{1+x}$Sn is one example \cite{douglas2014nanoscale, rice2017structural, douglas2014phase}. For such compounds, the stoichiometry is likely to only be self-limited to within the bounds of the phase field. However, for certain applications deliberately off-stoichiometric compositions are desired, e.g. the $\lambda_2=1$ criterion for low hysteresis shape memory alloys as described in the previous section \cite{james2000martensitic, james1989theory, bhattacharya1991wedge, ball100i987, gu2018phase}.

\textbf{Point defects.} Point defects in Heuslers also remain challenging to understand, measure, and control, in part because a quantitative experimental identification requires relatively low defect density samples. Our understanding is derived primarily from first-principles theory. DFT calculations on cubic full and half Heuslers predict a number of point defects, many with similarly small formation energies ($<1$ eV) \cite{yonggang2017natural, picozzi2007polarization, picozzi2004role}. The hexagonal polymorphs are less explored.

\begin{figure}[th]
 \includegraphics[width=3.5in]{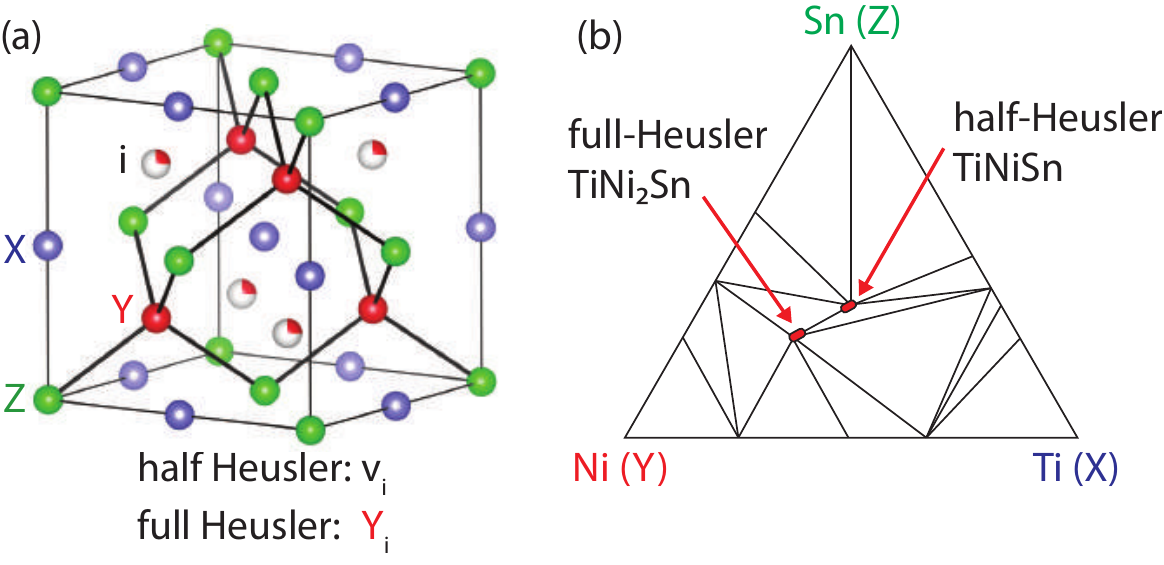}
 \caption{Defects and phase diagram for cubic Heuslers. (a) Crystal structure and defects. Half Heusler ($XYZ$) consists of an ordered vacancy sublattice $v_{i}$. A common defect for half Heuslers is a small fraction of $Y_{i}$ interstitials ($Y$ on interstitial $(\frac{3}{4}, \frac{1}{4}, \frac{1}{4})$ sites). The partially filled spheres denote fractional occupancy. In the limit of all vacancy sites being filled with $Y$, the structure is full Heusler ($XY_{2}Z$). For full Heuslers in which $X$ and $Z$ sites are indistinguishable, the structure is $B2$. (b) Ternary phase diagram of Ti-Ni-Sn at 497 $^{\circ}$C, adapted from Refs. \cite{romaka2013phase, stadnyk1991isothermal}. A tie line exists between the full and half Heusler phases. The phase fields for both TiNiSn and TiNi$_2$Sn are finite, and extend towards one another.}
 \label{phase}
\end{figure}

For half Heuslers, DFT calculations suggest the defect behavior may be grouped into families based on the chemical identity of the $Y$ site \cite{yonggang2017natural}. For $Y = 3d$ metal, $Y_{i}$ interstitials ($Y$ on interstitial $i$ sites, Fig. \ref{phase}a) are predicted to be the dominant low energy defect \cite{yonggang2017natural}, consistent with previous specific calculations for $Y=$ Ni \cite{larson2000structural, ogut}. These findings are consistent with the structural insight that $Y$ and $i$ sites have the same nearest neighbor coordination, and therefore filling these sites would have similar energies ($0$ to $0.5$ eV, depending on the position of the chemical potential \cite{yonggang2017natural}). In the dilute limit, $Y_i$ interstitials are expected to act as shallow donors \cite{yonggang2017natural}. In the high concentration limit they are expected to decrease the effective bandgap via the formation of an impurity band or a ``perturbed host'' band, which explains why many of the predicted semiconducting Heuslers behave experimentally as metals in transport measurements. The low formation energy for $Y_{i}$ is also proposed to drive a natural tendency for half Heuslers to be $Y$-rich \cite{yonggang2017natural}. This prediction of natural off stoichiometry is consistent with experimental observations that for TiNiSn, the phase field extends towards excess $Y=$ Ni and a thermodynamic tie line exists between half Heusler TiNiSn and full Heusler TiNi$_2$Sn (Fig. \ref{phase}b) \cite{douglas2014nanoscale, rice2017structural, douglas2014phase}. Such a tie line exists in many other half Heusler / full Heusler systems, e.g. ZrNiSn / ZrNi$_2$Sn and TiCoSn / TiCo$_2$Sn. Additionally, the electrical transport for TiNi$_{1+x}$Sn is optimized (low carrier density, high mobility) for samples that are slightly $Y=$ Ni rich ($x\sim 0.05$), suggesting that the excess Ni$_i$ compensates electrically for the natural Ni vacancy ($v_{Ni}$) formation \cite{rice2017structural}. A fruitful new direction for theorists would be to identify other electrically compensating point defects, to guide experimental efforts in making half Heuslers that are truly insulating.

For $Y= 5d$ half Heuslers, $Z_X$ antisites are expected to be the dominant defect, which act as acceptors \cite{yonggang2017natural}. General trends for $Y=4d$ are not well established \cite{yonggang2017natural}. 

Similar defect calculations for full Heuslers $XY_2 Z$ suggest $Y$ vacancies ($v_Y$) to be a low energy defect \cite{popescu2017native}, complementary to the $Y_i$ for half Heuslers. This prediction is consistent with experimental observations that full Heusler TiNi$_{2-x}$Sn has a finite phase field extending in the Ni-deficient direction \cite{romaka2013phase}. $X_Z$ and $Z_X$ antisites are another proposed defect in both full and half Heuslers \cite{larson2000structural, popescu2017native}, consistent with experimental observations of $B2$-type disorder \cite{khovailo2001order}, in which $X$ and $Z$ sites are indistinguishable, for films grown at low temperature \cite{rath2018reduced}. $X_Y$ and $Y_X$ antisites have also been proposed: in MnCo$_2$Si, first-principles calculations suggest Mn$_{Co}$ (Mn on Co lattice sites) have lowest formation energy and generally retain half metallic character, but other defects such as Co$_{Mn}$ are close in formation energy and can destroy half metallicity by forming states within the minority spin gap \cite{picozzi2004role}. Given the large zoology of proposed point defects for full Heuslers, many with similar small calculated formation energies ($<1$ eV \cite{picozzi2004role, picozzi2007polarization, yonggang2017natural}, compared to $\sim 3$ eV for self interstitials in Si \cite{rinke2009defect}), feedback between theory and measurements on clean samples are required to determine which defects are present, which are electronically active, and how to control them.

\textbf{Interdiffusion and reconstructions.} Most theoretical predictions assume idealized interfaces in which atoms adopt their bulk-like positions. However, at real materials interfaces there can be strong thermodynamic driving forces to deviate from simple bulk-like termination. This is especially important because interface properties are often inherently short-range. Heusler interfaces -- including Heusler/Heusler, Heusler/III-V, and Heusler/oxide -- are no exception. For Heuslers, the challenges exist at several length scales: interdiffusion and reactions at the several nanometer scale, and interfacial reconstructions and site disorder at the unit cell scale. 

Interdiffusion and interfacial reactions pose significant challenges at some Heusler/III-V semiconductor interfaces, particularly those containing Ni or Mn. This stems from the large diffusion coefficients for many transition metals in III-Vs ($D > 10^{-10}$ cm$^2$/s for Mn and Ni in GaAs at 500 $^\circ$C, compared to $D \sim 10^{-15}-10^{-13}$ for typical main group species \cite{wu1991diffusion, lahav1986interfacial, ruckman1986interdiffusion, fisher1998diffusion}), combined with complicated multi-component phase diagrams. These factors can result in interdiffused regions and secondary phases for direct Heusler on III-V growth at elevated temperatures ($>400^\circ$C \cite{gusenbauer2011interdiffusion}). Interdiffusion also limits the sharpness of Heusler/Heusler interfaces (e.g. MnCo$_2$Al / MnFe$_2$Al \cite{brown2018epitaxial}, TiNiSn / Zr$_{0.5}$Hf$_{0.5}$NiSn \cite{jaeger2014thermal}, MnNiSb /  MnPtSb \cite{mancoff1999growth}), but is generally less significant at Heusler/oxide interfaces due to the relative stability of many metal-oxides (FeCo$_2$Al / MgO \cite{bai2014magnetocrystalline}).

\begin{figure}[h]
 \includegraphics[width=3.5in]{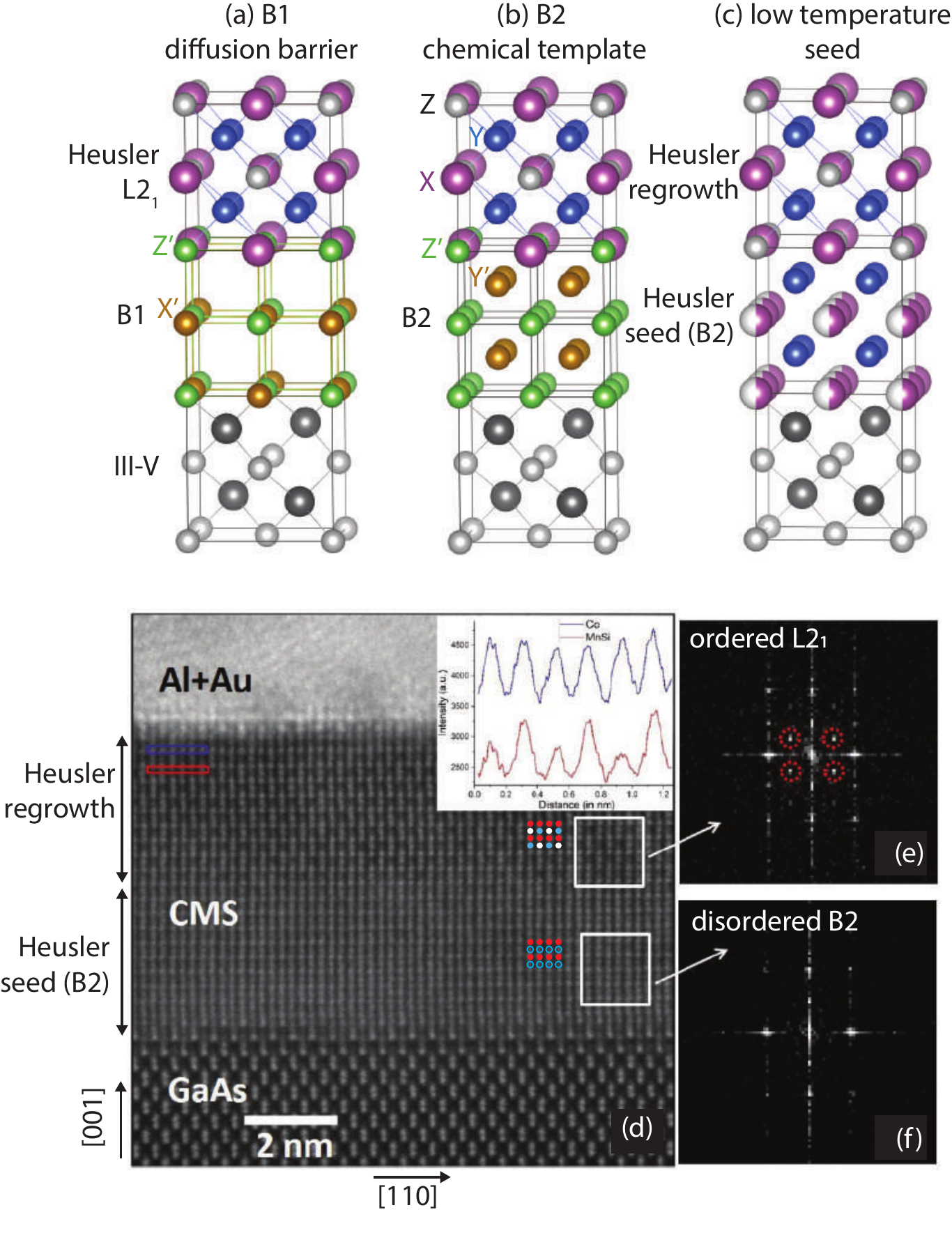}
 \caption{Strategies for making near atomically abrupt and stable interfaces. (a) Epitaxial $B1$ (rocksalt) diffusion barrier between Heusler film and and III-V substrate. One example diffusion barrier is ErAs, where $X'=$ Er and $Z'=$ As. (b) Epitaxial $B2$ (cesium chloride) chemical templating layer. Example NiGa, where $Y'=$ Ni and $Z'=$ Ga. (c) Low temperature seed layer, which minimizes interdiffusion but typically results in $B2$-type disorder at the interface. (d-f) Example of the low temperature seed layer approach. Reprinted figure with permission from A. Rath \textit{et. al.}, Phys. Rev. B 97, 045304 (2018) \cite{rath2018reduced}. Copyright 2002 by the American Physical Society. (d)  Cross sectional high angle annular dark field - scanning trasmission electron microscopy (HAADF-STEM) image of the interface between MnCo$_2$Si (also known as Co$_2$MnSi, CMS) and GaAs (001). Within 5 nm of the interface the CMS seed has disordered $B2$ structure, while the top region shows the fully ordered $L2_1$ structure. For this sample the growth temperature was held constant at $270 ^\circ$C. (e,f) Fast Fourier transform of the regrowth and seed regions.}
 \label{diffusion}
\end{figure}

One solution is to grow epitaxial diffusion barriers between the Heusler film and III-V substrate (Fig. \ref{diffusion}a). The rare earth monopnictides ($RV$, $R=$ rare earth, $V=$ As, Sb, Bi) are highly effective diffusion barriers for group III and transition metal species \cite{schultz2002eras, palmstrom1993stable}. These materials have cubic rocksalt ($B1$) structure and can can be lattice matched by alloying on the rare earth site. Examples include ErAs, Sc$_x$Er$_{1-x}$As, ErSb, and GdSb, which have enabled the epitaxial growth of a variety of intermetallic films on III-Vs at temperatures up to $600^\circ$C \cite{schultz2002eras, buschbeck2011growth, palmstrom2016heusler, palmstrom1993stable}. However, the rare earth monopnictides are generally metallic, magnetic, and require a finite thickness of at least three atomic layers to be effective diffusion barriers. Hence they are not suitable when a direct Heusler/III-V interface is required. 

Another approach is to grow thermodynamically stable, chemical templating layers \cite{palmstrom1993stable, filippou2018chiral} (Fig. \ref{diffusion}b). $B2$ interlayers (cesium chloride structure) are good templates for full Heusler growth, since these two structures are ordered variants of one another. Starting from the cubic $B2$ structure, whose basis consists of $Z (Z')$ at the origin and $Y (Y')$ at the body center, the full Heusler $L2_1$ structure is obtained by replacing every other $Z$ site with $X$. One example is to use a $B2$ NiGa interlayer to seed the growth of MnNi$_2$Ga on GaAs (001). NiGa is thermodynamically stable in contact with GaAs \cite{dong2000epitaxial}, thus minimizing the interdiffusion. $B2$ templating can also enhance the $c$-axis ordering in Heuslers \cite{filippou2018chiral}, since the [001] stacking sequence of a $B2$ crystal with composition $Y'Z'$ consists of alternating atomic planes of $Y'$ and $Z'$. This template enhances the $c$ axis ordering of the subsequent Heusler film, due to the local bonding preference of $Y$ on $Z'$ and $XZ$ on $Y'$. However, like the rocksalt $B1$ diffusion barriers, $B2$ template layers are often metallic and require a finite thickness, and are also not suitable when a direct Heusler/III-V interface is required.

For direct Heusler/III-V interfaces or for interfaces between two different Heusler compounds, low temperature seed layers are the method of choice \cite{palmstrom2016heusler, farshchi2013spin} (Fig. \ref{diffusion}c). This strategy consists of nucleating several unit cells of Heusler film at low temperature ($< 300 ^\circ$C) to minimize interdiffusion during the formation of the interface \cite{hashimoto2007atomic, kawasaki_cotisb}. The seed can be then be annealed and growth resumed at higher temperatures \textbf{($\sim 500^\circ$C)} to improve the degree of $L2_1$ ordering \cite{kawasaki_cotisb, hirohata2005structural, hashimoto2007atomic, farshchi2013spin, palmstrom2016heusler}. This strategy relies on the fact that bulk diffusion is generally much slower than surface diffusion during growth. Once the interface is formed at low temperature, interdiffusion is suppressed for subsequent anneals compared to direct growth at higher temperatures, as inferred by reflection high energy electron diffraction and x-ray diffraction \cite{hirohata2005structural, kawasaki_cotisb} or by device performance metrics such as the resistance-area product of a magnetoresistance junction \cite{kubota2017current}. Direct measurements of the interdiffusion, e.g. by Rutherford Backscattering Spectrometry or STEM-EELS, are needed to fully quantify these effects as a function of post-growth anneal temperature. 

For Heusler/III-V \cite{rath2018reduced, nedelkoski2016realisation} and Heusler/Heusler \cite{brown2018epitaxial} interfaces formed by low temperature seeds, the chemical intermixing is typically limited to a few atomic layers (Fig. \ref{diffusion}d) \cite{rath2018reduced}. However, due to the low surface diffusion at low temperatures, the seed layers often crystallizes in the disordered $B2$ structure, in which $X$ and $Z$ sites are indistinguishable, rather than the ordered full Heusler $L2_1$ \cite{farshchi2013spin, rath2018reduced}. The effects of such disorder on properties can vary significantly depending on the particular compound and desired property \cite{orgassa2000disorder,farshchi2013spin, palmstrom2016heusler, wang2005magnetic,inomata2008site}. Low temperature growth also impedes the ability to control stoichiometry and point defects, which are better controlled under high temperature, adsorption-controlled growth regimes \cite{kawasaki2018simple, patel2014surface, bach, turban, strohbeen2019electronically}. 

Even for highly controlled chemical abruptness, thermodynamic driving forces can cause interfacial layer relaxations, atomic reconstructions, and even layer rearrangements. An extreme example is the MnCo$_2$Si / GaAs (001) interface. The bulk (001) atomic stacking sequence of MnCo$_2$Si (also known as Co$_2$MnSi) consists of alternating atomic layers of MnSi and CoCo; however, photoemission spectroscopy measurements reveal that this interface tends to be Mn and As-rich, independent of whether the MnCo$_2$Si growth on As-terminated GaAs is initiated with a monolayer of MnSi or CoCo \cite{palmstrom2016heusler, patelthesis}. Such atomic layer rearrangements are not unique to Heuslers, for example, they are also observed in layered perovskite oxides \cite{nie2014atomically, lee2014dynamic}. 

The strong thermodynamic driving forces place constraints on what interfaces can be synthesized, which is an important consideration since interface electronic states, half metallicity, charge transfer, and other interfacial properties can be highly sensitive to the interface termination \cite{picozzi2007polarization}. Feedback from theory is crucial for identifying which types of interfaces are both stable and host the desired property \cite{curtarolo2003predicting, sun2016thermodynamic, curtarolo2005accuracy, zunger2018realization}. A significant challenge is that interfaces have reduced symmetry and increased atomic degrees of freedom, compared to the bulk. Given this potential complexity, it often is not practical to perform first-principles calculations for all possible interface atomic structures. There are too many candidate structures, and the large size of reconstructed slabs makes first-principles approaches computationally expensive. Simple models based on electron counting have recently been developed to guide the screening of stable structures at surfaces \cite{kawasaki2018simple}, which can be down selected for more accurate first-principles calculations. I anticipate that their generalization may make the interface problem more tractable.

\section*{Outlook}


Heusler compounds are a remarkable family of interfacial materials, whose broad range of tunable properties is highly complementary to that of the well-studied transition metal oxides. These compounds are lattice-matched to technologically important semiconductor substrates, making them poised for impact in spintronics and beyond. I conclude with a few remarks on the role of theory and experiments going forward.

\textbf{Theory.} To date, theory has done an excellent job at screening for target properties in the bulk \cite{oliynyk2016high, carrete2014finding, garrity2014pseudopotentials, bennett2012hexagonal, garrity2014hyperferroelectrics, roy2012half, anand2019enormous, carrete2014finding, sanvito2017accelerated} and predicting emergent properties at idealized interfaces, both at the level of first-principles DFT calculations \cite{chadov2010tunable, lin2010half, picozzi2007polarization, picozzi2004role, zhu2015surface, narayan2015class} and model Hamiltonians \cite{timm2017inflated, brydon2016pairing, venderbos2018pairing}. Can such predictions be modified to account for more realistic structural distortions at Heusler interfaces, including relaxations, reconstructions, and point defects? Additionally, can theory aid in identifying which of these compounds and interfaces are thermodynamically stable, or more relevantly, ``stabilizable?'' New theoretical approaches are beginning to consider the path-dependent ``remnant'' metastability of bulk compounds \cite{sun2016thermodynamic}, to identify which compounds have local minima in the free energy landscape that lie not too far above the convex hull \cite{aykol2018thermodynamic, curtarolo2003predicting, curtarolo2005accuracy, zunger2018realization}, and guide possible synthesis routes \cite{sun2016thermodynamic, chen2018understanding}. To what extent can these concepts be applied to Heuslers, and in particular, Heusler interfaces?

\textbf{Experiments.} Heusler compounds today are comparable to semiconductors in the early 20$^{th}$ century. Although field effect transistors were first proposed in the 1920's and 30's \cite{edgar1930method, edgar1933device, heil1935improvements}, the first experimental demonstrations of point contact transistors \cite{PhysRev.74.230} and field effect transistors \cite{arns1998other, dawon1963electric} were not made until the late 40's and 50's. These discoveries were made possible by two major materials and interface innovations: (1) zone refining of germanium and silicon to reduce the background impurities, and (2) methods to prepare clean semiconductor / oxide interfaces, free of trapped charges. 

Heusler compounds today are at a similar stage of development: a number of exotic phenomena have been predicted, but their full realization will likely require new advances in materials synthesis and interface control. In this Perspective I outlined a few of the key synthetic challenges and potential solutions. I look forward to the development of new feedback control methods during growth, new chemical precursors for self-limiting stoichiometry, and new methods to probe the properties of buried interfaces. Beyond the significant advances in Heusler spintronics, the broader field of Heusler interfaces is at a stage of relative infancy. I anticipate that the most exotic and the impactful properties of Heusler interfaces have yet to be unleashed.


\section{Acknowledgments}

I thank Chris J. Palmstr{\o}m for insightful feedback throughout the preparation of this manuscript and for his mentorship. I also thank Darrell G. Schlom and Jochen Mannhart \cite{mannhart2010oxide} for inspiring the title. I thank Richard D. James, Anderson Janotti, Chang-Beom Eom, Darrell G. Schlom, and Paul M. Voyles for their feedback, fruitful discussions, and collaborations.

This work was primarily supported by the CAREER program of the National Science Foundation (NSF DMR-1752797) and by the Young Investigator Program of the Army Research Office (ARO W911NF-17-1-0254). Additional support came from the NSF via the University of Wisconsin Materials Research Science and Engineering Center (MRSEC, DMR-1720415) and the Wisconsin Alumni Research Foundation (WARF).

\bibliographystyle{apsrev}

\bibliography{bibliography}

\end{document}